\documentclass[12pt]{iopart}
\usepackage{graphicx}
\usepackage{iopams}

\renewcommand{\o}{\textrm{\scriptsize o}}
\renewcommand{\b}{\textrm{\scriptsize b}}
\newcommand{\LS}{\textrm{\scriptsize LS}}
\newcommand{\I}{\textrm{\scriptsize I}}
\renewcommand{\l}{\left<}
\renewcommand{\r}{\right>}

\begin{document}
\title{Variational principle for bifurcation in Lagrangian mechanics}
\author{
Toshiaki Fujiwara$^1$,
Hiroshi Fukuda$^1$
and Hiroshi Ozaki$^2$}

\address{$^1$ College of Liberal Arts and Sciences Kitasato University, 1-15-1 Kitasato, Sagamihara, Kanagawa 252-0329, Japan}
\address{$^2$ Laboratory of general education for science and technology, Faculty of Science, Tokai University, 4-1-1 Kita-Kaname, Hiratsuka, Kanagawa, 259-1292, Japan}
\ead{fujiwara@kitasato-u.ac.jp,
fukuda@kitasato-u.ac.jp
and ozaki@tokai-u.jp}

\begin{abstract}
An application of variational principle
to bifurcation of periodic solution in Lagrangian mechanics is shown.
A few higher derivatives of the action integral at a periodic solution
reveals the behaviour of the action in function space 
near the solution.
Then the variational principle gives a method to find
bifurcations from the solution.
The second derivative (Hessian) of the action
has an important role.
At a bifurcation point, an eigenvalue of Hessian tends to zero.
Inversely, if an eigenvalue tends to zero, the zero point is a bifurcation point.
The third and higher derivatives of the action
determine the properties of the bifurcation
and bifurcated solution.
\end{abstract}
\vspace{2pc}



\section{Variational principle}
Variational principle demands that
a function that gives stationary point of the action integral is
a solution of motion \cite{Landau,Arnold}.
In this paper, we consider dynamical system described by
the following Lagrangian $L$ and the action integral $S$
with $N$ degrees of freedom,
\begin{equation}
\label{LagrangianAndAction}
L= \sum_{k=1}^N \frac{m_k}{2}\left(\frac{d q_k}{d t}\right)^2
			+V(q_1,q_2,\dots,q_N),\qquad
S=\int d t L.
\end{equation}
The variables $q_k$ can be generalized coordinates.
However, we assume the kinetic term is the quadratic form of $d q_k/d t$,
and $m_k$ are constants.
The variational principle produces the equation of motion,
\begin{equation}
\delta S
=0
\qquad\Leftrightarrow\qquad
-m_k\frac{d^2 q_k}{d t^2}+\frac{\partial V}{\partial q_k}=0.
\end{equation}

The variational method to find a solution of the equation of motion
is an application of the variational principle.
Namely, a minimum or maximum point of the action $S$
in function space 
is a solution of the equation of motion,
since minimum or maximum satisfies $\delta S=0$ \cite{Moore}.
A proof of existence of
minimum  or maximum 
of the action
gives a proof of existence of a solution \cite{ChencinerMontgomery}.

The aim of this paper is to describe a method to find  
bifurcation point and bifurcated function
emerged from a periodic solution based on the variational principle.
Suppose we get a few derivatives of the action at a solution,
$\delta^2 S, \delta^3 S, \dots$,
we will see the values of the action varies with place to place in functional space around the solution.
Then, if there is a  stationary point other than the original solution,
the point must be a new solution by the principle.
Since, with a few derivatives,
the region we can see around the original solution is very restricted,
the other stationary point is usually out of the region.
Fortunately,
observation near the bifurcation point
will give us a chance to find the original and other stationary point.
Suppose the Lagrangian has a parameter $\xi$,
and $\xi=\xi_0$
is a bifurcation point.
If we observe  the action in a range of $\xi$ that include $\xi_0$,
we will see the bifurcated solution come into our region,
and finally reaches to the original solution at $\xi_0$.
Thus, bifurcations can be described by  a few derivatives of the action.

How do we find the bifurcation point
observing only the original solution?
In other words, how do we know that another stationary point
is approaching to the original?
If the action $S$ were a function of single variable $x$,
approaching two stationary points that satisfy
$S'(x_\o)=S'(x_\b)=0$
makes the second derivative $S''(x_\o)$ zero
for the limit $x_\b \to x_\o$.
Indeed, in section~\ref{secSecondDerivative},
we will show that
the second derivative of the action, described by Hessian operator,
must have zero eigenvalue at the bifurcation point.
Let $\kappa$ be the eigenvalue that tends to zero,
\begin{equation}
\label{eqTheLimit}
\kappa \to 0 \mbox{ for } \xi \to \xi_0.
\end{equation}
This is a necessary condition for bifurcation.
There are two possibilities for \eref{eqTheLimit},
\begin{enumerate}
\item	$\kappa\ne 0$ and $\kappa \to 0$ for $\xi \to \xi_0$,\label{theMainCase}
\item	$\kappa$ is always zero.\label{conservedCase}
\end{enumerate}
In this section, we just consider the case (\ref{theMainCase}).
The case (\ref{conservedCase}) is irrelevant to bifurcations, that will be discussed later.

The answer of the question is the following,
if an eigenvalue of the Hessian of the original solution tend to zero,
a bifurcation may appear.
Inversely, if no eigenvalue tends to zero, no bifurcation appears.

What will happen if $\kappa\ne 0$ and $\kappa\to 0$.
Again, if the action were a function of single variable $x$,
for example,
\begin{equation}
\label{eqThiredOrder}
S(x)=S(0)+\frac{\kappa}{2} x^2 - \frac{A_3}{3!} x^3 +\Or(x^4),\qquad A_3\ne0,
\end{equation}
then the condition for stationary points
\begin{equation}
S'(x)=x\left(\kappa-\frac{A_3}{2}x+\Or(x^2)\right)=0
\end{equation}
has two zeros.
The solution $x_\o=0$ stands for the original solution
and 
\begin{equation}
x_\b=\frac{2}{A_3}\kappa + \Or(\kappa^2)
\end{equation}
for the bifurcated solution.
Thus 
if the second derivative of the original solution $S''(0)=\kappa$
tends to zero,
a bifurcation appears.
We now know in the case \eref{eqThiredOrder} that
the bifurcated solution exists both sides of $\kappa>0$ and $\kappa<0$.
We also know that the action and the second derivative of the action
at the bifurcated solution are
\begin{equation}
\label{eqValueOfActionAndEigenValueOfHessian}
S(x_\b)=S(x_\o)+\frac{2}{3 A_3^2}\kappa^3+\Or(\kappa^4),\qquad
S''(x_\b)=-\kappa+\Or(\kappa^2).
\end{equation}

Although the action is a function of infinitely many variables,
Lyapunov-Schmidt reduction makes the action 
to be a function of one or a few variables.
This procedure will be shown  in section~\ref{reducedAction}.
Using the reduced action,
we will show that if an eigenvalue of Hessian tends to zero, 
the point zero is a bifurcation point.
This gives a sufficient condition for bifurcation.

The Lyapunov-Schmidt reduced action also describes
in which side of $\kappa$ the bifurcated solution exists,
and describes value of the action 
and the eigenvalue of Hessian at the bifurcated solution
as shown in \eref{eqValueOfActionAndEigenValueOfHessian}.
It will be shown in section~\ref{secThirdOrder} and \ref{secFourthOrder}.

Summary and discussions are given in section~\ref{secDiscussions}.
Applications of this method to
individual periodic solutions in Lagrangian mechanics
will be published separately.

\section{The second derivative of the action}
	\label{secSecondDerivative}
In this section,
we first define Hessian operator that describes the second derivative of the action.
Then, we describe the necessary condition for bifurcations.

\subsection{Definition of Hessian}
Now, let us calculate the second derivative of the action~(\ref{LagrangianAndAction}) at 
a periodic solution $q(t)$,
\begin{equation}
S(q+\delta q)
=S(q)+\frac{1}{2}\int_0^T d t \sum_{i j}\delta q_i\left(
	-m_i \delta_{i j} \frac{d^2}{d t^2}
	+\frac{\partial^2 V}{\partial q_i \partial q_j}
	\right)
	\delta q_j
+\Or(\delta q^3),
\end{equation}
where the symbol $\delta_{i j}$ is the Kronecker delta,
and
\begin{equation}
q(t+T)=q(t),\qquad \delta q(t+T)=\delta q(t).
\end{equation}
We restricted the function space to be periodic
with period $T$ which is the same as that of the solution.
The first order of $\delta q$ is zero, since $q$ is a solution of the equation of motion.

The second derivative defines Hessian operator $\mathcal{H}$,
\begin{equation}
\mathcal{H}_{i j}=-m_i \delta_{i j} \frac{d^2}{d t^2}
	+\frac{\partial^2 V}{\partial q_i \partial q_j}.
\end{equation}
Hereafter, we use the vector and matrix notation,
$m=(m_i \delta_{i j})$, $q=(q_i)$, 
$V_q=(\partial V/\partial q_i)$, and $\mathcal{H}=(\mathcal{H}_{i j})$
with $i, j=1,2,\dots,N$.

Then, the action up to the second order is
\begin{equation}
\fl
S(q+\delta q)=S(q)+\frac{1}{2}\int_0^T d t\ \delta q \mathcal{H}\delta q + \Or(\delta q^3),
\qquad
\mathcal{H}=-m\frac{d^2}{d t^2}+V_{q q}
\end{equation}
and the equation of motion is
\begin{equation}
m\frac{d^2 q}{d t^2}=V_q(q).
\end{equation}

\subsection{Necessary condition for bifurcation}
	\label{secNecessaryCondition}
Let us assume the Lagrangian contains a parameter $\xi$,
and at $\xi=\xi_0$ the solution $q_\o$ bifurcates to make 
a bifurcated solution $q_\b$.
The period $T$ may or may not depend on the parameter $\xi$.
At $\xi$ being very close to $\xi_0$
where two solutions exists,
the following equations of motion are satisfied,
\begin{equation}
\label{eqForOriginalAndBifurcated}
m\frac{d^2 q_\b}{d t^2}=V_q(q_\b),
\qquad
m\frac{d^2 q_\o}{d t^2}=V_q(q_\o).
\end{equation}
Let the difference of the two solutions be 
$q_\b-q_\o=R \Phi$, $R\ne 0$ and $\int_0^T d t \Phi^2=1$.
Difference of  two equations yields
\begin{equation}
\label{eqForOriginalAndBifurcated2}
R  m\frac{d^2 \Phi}{d t^2}
=V_q(q_\o+R\Phi)-V_q(q_\o)
=R V_{q q}\Phi + \Or(R^2).
\end{equation}
Dividing by $R$, we obtain
\begin{equation}
\label{eqForOriginalAndBifurcatedHessian}
\mathcal{H}\Phi=\Or(R).
\end{equation}
The limit $\xi \to \xi_0$ makes $R \to 0$ and 
\begin{equation}
\label{zeroEigenvalue}
\mathcal{H}\Phi \to 0.
\end{equation}
This is a necessary condition for bifurcation.
Namely, at the bifurcation point,
the Hessian must have zero eigenvalue.

Now, we expand the difference $q_\b-q_o$
by the eigenfunctions of Hessian,
\begin{equation}
q_\b-q_o=r\phi+r\sum_\alpha \epsilon_\alpha \psi_\alpha,
\qquad
\mathcal{H}\phi=\kappa \phi,
\qquad
\mathcal{H}\psi_\alpha=\lambda_\alpha \psi_\alpha,
\end{equation}
and consider the contents of  the condition \eref{zeroEigenvalue}.
We use Greek indexes $\alpha, \beta, \dots$ to distinguish functions,
while roman $i, j, \dots$ to express the vector component of a function.
The functions $\phi$ and $\psi_\alpha$ are normalized orthogonal functions
\begin{equation}
\int_0^T dt\, \phi^2 =1,
\int_0^T dt\, \phi_\alpha \psi_\beta = \delta_{\alpha\beta}
\mbox{ and } \int_0^T dt\, \phi \psi_\alpha=0.
\end{equation}
The function $\phi$ is the eigenfunction
that is shown to exist by the condition \eref{zeroEigenvalue}.
Namely $\kappa \to 0$ for $\xi \to \xi_0$.
The same arguments for \eref{eqForOriginalAndBifurcated}
to \eref{eqForOriginalAndBifurcated2}
yields
\begin{equation}
r \mathcal{H}\, (\phi+\sum_\alpha \epsilon_\alpha \psi_\alpha)
=r  (\kappa \phi+\sum_\alpha \epsilon_\alpha \lambda_\alpha \psi_\alpha)
=O(r^2).
\end{equation}
Dividing by $r\ne 0$, we obtain 
\begin{equation}
\kappa \phi+\sum_\alpha \epsilon_\alpha \lambda_\alpha \psi_\alpha=O(r).
\end{equation}
Let us take a small region of $\xi$ that contains $\xi_0$,
where 
only $\kappa \to 0$ for $\xi \to \xi_0$
and all $|\lambda_\alpha|$ are grater than a positive number.
(We exclude zero eigenvalues that are always zero for 
any parameter $\xi$. Discussion for  zero eigenvalues will be given soon later.)
Since, $\phi$ and $\psi_\alpha$ are orthogonal normalized functions
and $|\lambda_\alpha|$ is grater than a positive number,
\begin{equation}
\label{necessaryCondition2}
\kappa = O(r),\ \epsilon_\alpha = O(r)
\mbox{ for } r\to 0.
\end{equation}
This is the contents of \eref{zeroEigenvalue}
for necessary condition for bifurcation.
Since the difference
$q_\b-q_o=r\phi+r\sum \epsilon_\alpha \psi_\alpha$
and $r\epsilon_\alpha = O(r^2)$,
the function $\phi$ is the primary part and $\psi_\alpha$ are secondary part
for the difference.

Zero eigenvalues that always exists for any values of the  parameter $\xi$
are irrelevant to any bifurcations.
They are connected to the conservation laws.
For example,
the energy conservation law
is connected to a shift of the origin of time, $q(t) \to q(t+r)$
by Noether's theorem \cite{Landau,Arnold}.
The corresponding eigenfunction is proportional to $d q_\o/d t$.
Actually, the action is flat along the curve
parametrized by $r$,
\begin{equation}
\delta q
=q_\o(t+r)-q_\o(t)
=r \frac{d q_\o}{d t} +\sum_{n\ge 2} \frac{r^n}{n!}\frac{d^n q_\o(t)}{d t^n}.
\end{equation}
This eigenvector $d q_\o/d t$ does not make bifurcation.
Moreover, 
we can exclude $d q_\o/d t$ from the set $\{\psi_\alpha\}$ by fixing the origin of time.
In general, eigenfunctions that belong to zero eigenvalues
are irrelevant to bifurcations 
and can be excluded from the set $\{\phi, \psi_\alpha\}$.

In the next section, 
we will show that
if an eigenvalue $\kappa\ne 0$ tends to zero,
the zero point is a bifurcation point.

\section{Lyapunov-Schmidt reduction of the action
and a sufficient condition for bifurcation}\label{reducedAction}

In this section, we will show that the action
$S(q_\o + r\phi+r\sum \epsilon_\alpha \psi_\alpha)$
actually has stationary point
if $\kappa\ne 0$ tends to zero.
This gives the sufficient condition for bifurcation.

The eigenvalue $\kappa$ may be degenerated.
We first consider non-degenerated case in subsection \ref{secNondegenerate}
and \ref{secSufficientCondition}.
Degenerated case will be considered in \ref{secDegenerate}.

\subsection{Reduced action for non-degenerated case}
\label{secNondegenerate}
In this subsection, we assume the eigenvalue $\kappa$ is not degenerated.
So, the eigenfunction $\phi$ belongs to $\kappa$ is unique.
The action $S(q_\o+r\phi+r\sum \epsilon_\alpha \psi_\alpha)$
is given by the expansion series of 
$r$ and $\epsilon_\alpha$,
\begin{eqnarray}
\label{expansionOfS}
S(r, \epsilon)
&=S(q_\o+r\phi+r\sum \epsilon_\alpha \psi_\alpha)\nonumber\\
&=S(q_\o)
	+\frac{\kappa}{2}r^2
	+\sum \frac{\lambda_\alpha}{2}\epsilon_\alpha^2 r^2
	+\sum_{n\ge 3}\frac{r^n}{n!} \l \phi^n \r\nonumber\\
	&+\sum \epsilon_\alpha \sum_{n\ge 2}\frac{r^{n+1}}{n!} \l \phi^n \psi_\alpha \r
	+\sum \frac{\epsilon_\alpha \epsilon_\beta}{2} \sum_{n\ge 1}\frac{r^{n+2}}{n!} \l \phi^n \psi_\alpha \psi_\beta \r
	+\dots,
\end{eqnarray}
where for functions $f_1, f_2, \dots, f_n$,
\begin{equation}
\l f_1 f_2 \dots f_n \r
=\int_0^T d t\ \frac{\partial^n V}{\partial q^n}f_1 f_2 \dots f_n.
\end{equation}
The vector indexes of functions are  understood to be properly contracted
with the indexes of derivative of $V$,
for example,
\begin{equation}
\l fg\r
=\int d t\ \frac{\partial^2 V}{\partial q_i \partial q_j} f_i g_j,\ 
\l f^2\r
=\int d t\ \frac{\partial^2 V}{\partial q_i \partial q_j} f_i f_j,\ 
\dots.
\end{equation}

By the variational principle,
the solutions of stationary conditions
\begin{equation}
\label{eqFullForBifurcation}
\partial_r S(r,\epsilon)=\partial_{\epsilon_\alpha} S(r, \epsilon)=0
\end{equation}
are solutions of equation of motion.
From among them, 
we choose  solutions that satisfy
$\epsilon_\alpha \to 0$ and $\kappa \to 0$ for $r \to 0$,
since we are looking for bifurcated solutions.

Instead of solving the conditions \eref{eqFullForBifurcation} at once,
we first solve $\epsilon_\alpha(r)$ for arbitrary $r$ and put them into the action
to define the  reduced action,
\begin{equation}
S_\LS(r)=S(r, \epsilon(r)).
\end{equation}
This process is known as Lyapunov-Schmidt reduction  \cite{RotatingEight, GuoWu2013}.
Since
\begin{equation}
S'_\LS(r)
=\frac{d S_\LS(r)}{d r}
=\partial_r S(r,\epsilon)
	+\sum_\alpha
		\left.
		\partial_{\epsilon_\alpha}S(r,\epsilon)
		\right|_{\epsilon=\epsilon_\alpha(r)}\frac{d \epsilon_\alpha(r)}{d r},
\end{equation}
the condition $S'_\LS(r)=0$ with $\epsilon_\alpha=\epsilon_\alpha(r)$
is equivalent to the conditions \eref{eqFullForBifurcation}.
Therefore, if $S'_\LS(r)=0$ has solution $r=r_\b \ne 0$ and $r_\b \to 0$ for $\kappa \to 0$,
a bifurcated solution
$q_\b=q_\o+r_\b\phi+r_\b \sum \epsilon_\alpha(r_\b) \psi_\alpha$ exists.
The value of the action is also correctly given by the reduced action, $S_\LS(r_\b)=S(r_\b, \epsilon(r_\b))$.

Now, let us solve the condition \eref{eqFullForBifurcation} for $\epsilon$.
\begin{equation}
\fl
\partial_{\epsilon_\alpha} S(r,\epsilon)
=r^2\left(
	\lambda_\alpha \epsilon_\alpha
	+\sum_{n\ge 2}\frac{r^{n-1}}{n!}\l \phi^n \psi_\alpha \r
	+\sum_\beta \epsilon_\beta \sum_{m\ge 1}\frac{r^m}{m!}\l \phi^m \psi_\alpha \psi_\beta \r
	+\dots
	\right)
=0.
\end{equation}
The solution that satisfy $\epsilon_\alpha \to 0$ for $r \to 0$ is
uniquely determined recursively,
\begin{equation}
\label{solutionOfEpsilon}
\epsilon_\alpha
=-\sum_{n\ge 2}\frac{r^{n-1}}{n!}\frac{\l \phi^n \psi_\alpha\r}{\lambda_\alpha}
	+\sum_{m \ge 1, n\ge 2}\frac{r^{m+n-1}}{m! n!}
	\sum_\beta \frac{\l\phi^m\psi_\alpha \psi_\beta\r \l \phi^n \psi_\beta\r}{\lambda_\alpha 
	\lambda_\beta}
	+\dots
\end{equation}
The lowest order term is
\begin{equation}
\epsilon_\alpha = -\frac{r}{2\lambda_\alpha}\l \phi^2 \psi_\alpha \r + \Or(r^2).
\end{equation}
Substituting this solution into \eref{expansionOfS}, we obtain the reduced action
\begin{eqnarray}
\label{eqSreducedForOrder4}
\fl S_\LS(r)
&=S(q_\o)+\frac{\kappa}{2}r^2
	+\frac{r^3}{3!}\l \phi^3 \r
	+r^4\left(\frac{1}{4!}\l \phi^4 \r -\sum\frac{1}{8\lambda_\alpha}\l \phi^2 \psi_\alpha \r^2\right)
	+\dots\nonumber\\
\fl &=S(q_\o)+\frac{\kappa}{2}r^2 -\frac{A_3}{3!}r^3 - \frac{A_4}{4!}r^4 + \Or(r^5), 
\end{eqnarray}
where
\begin{equation}
A_3=-\l \phi^3 \r,\qquad
A_4=-\l \phi^4 \r +\sum_\alpha \frac{3}{\lambda_\alpha}\l \phi^2 \psi_\alpha \r^2,\qquad
\dots.
\end{equation}

Thus we obtain the reduced action $S_\LS(r)$ that depends on single variable $r$.
Other variables $\epsilon_\alpha$ are eliminated by the condition for stationary point .

\subsection{Sufficient condition for bifurcation}
\label{secSufficientCondition}
Let the reduced action generally be
\begin{equation}
S_\LS(r)=S(q_\o)+\frac{\kappa}{2}r^2  -\sum_{n\ge 3}\frac{A_{n}}{n!}r^n.
\end{equation}
Then the condition for stationary point is
\begin{equation}
S'_\LS(r)
=r\left(\kappa-\sum_{n\ge 3}\frac{A_n}{(n-1)!}r^{n-2}\right)
=0.
\end{equation}
the solution $r_\o=0$ stands for the original solution,
and the solution of
\begin{equation}
\label{solutionOfKappa}
\kappa=\sum_{n\ge 3}\frac{A_n}{(n-1)!}r_\b^{n-2}
\end{equation}
stands for a bifurcation,
Then the action at the bifurcated solution is
\begin{equation}
S(r_\b)=S(q_\o)+\sum_{n\ge 3}\frac{(n-2)}{2}\frac{A_n}{n!}r_b^n.
\end{equation}
%
So far, we have shown that there are bifurcated solution
given in series expansion of $r$, namely
\eref{solutionOfEpsilon} and \eref{solutionOfKappa},
which certainly satisfies conditions \eref{necessaryCondition2}.

Now, we proceed to the last step for  sufficient condition
by solving the series \eref{solutionOfKappa} inversely.
If $A_3\ne 0$, a bifurcated solution
\begin{equation}
r_\b=\frac{2}{A_3}\kappa +\Or(\kappa^2)
\end{equation}
appears at $\kappa=0$.
We call this a third order bifurcation.
Whereas, if $A_3=0$ and $A_4\ne 0$, a bifurcated solution
\begin{equation}
r_\b=\pm \left(\frac{6\kappa}{A_4}\right)^{1/2}(1+\Or(\kappa))
\end{equation}
appears. We call this a fourth order bifurcation.
In general, 
if $A_3=A_4=\dots=A_{n-1}=0$ and $A_n\ne 0$, a bifurcation
\begin{equation}
r_\b=\left((n-1)! \frac{\kappa}{A_n}\right)^{1/(n-2)}(1+\Or(\kappa))
\end{equation}
will appear, that we call a $n$-th order bifurcation.
The bifurcated solution $r_\b$ exists both $\kappa<0$ and $\kappa>0$ for odd $n$.
Whereas it exists only one side of $\kappa$ for even $n$,
$\kappa>0$ side if $A_n>0$ and $\kappa<0$ side if $A_n<0$.
Therefore, for almost all cases, a bifurcation appears at $\kappa=0$.
So, $\kappa \ne 0$ for $\xi\ne \xi_0$ and $\kappa \to 0$ for $\xi \to \xi_0$
is the sufficient condition for bifurcation.

One possible  exception is the case $A_n=0$ for all $n$.
In this case, however,
the reduced action is exactly
$S_\LS(r)=S(q_\o)+\kappa r^2/2$.
This action behaves badly.
Consider the behaviour  of this reduced action at sufficiently large $r=M$.
For the small interval of $-1/M < \kappa < 1/M$,
the change of action is huge, since $-M/2 < \kappa r^2/2 <M/2$.
Although, we didn't find a logic to exclude this case,
this case unlikely exists. 

Before leaving this subsection,
let us calculate the distance between the bifurcated and the original solution
that is defined by,
\begin{equation}
R^2=
||q_\b-q_\o||^2=\int_0^T d t (q_\b-q_\o)^2
=r^2\left(1+\sum_\alpha \epsilon_\alpha^2\right)
=r^2 + \Or(r^4),
\end{equation}
because $\epsilon_\alpha^2=\Or(r^2)$.

\subsection{Degenerated case}
\label{secDegenerate}
Now, let us consider the case $\kappa$ is degenerated
with degeneracy number $g$.
In this case, there are $g$ linearly independent eigenfunctions
for $\mathcal{H}\phi_\gamma = \kappa \phi_\gamma$, $\gamma=1,2,\dots, g$.
In this case,
any function in  the space spanned by $\phi_\gamma$ can be expressed
as $r\phi(\theta)$ using a polar coordinates
$(r, \theta)=(r, \theta_1, \theta_2, \dots, \theta_{g-1})$.
For example, $r\phi(\theta)=r(\cos(\theta)\phi_1+\sin(\theta)\phi_2)$ if $g=2$.

Following the same arguments in \sref{secNondegenerate},
we obtain the reduced action $S_\LS(r,\theta)$,
\begin{equation}
S_\LS(r,\theta)
=S(q_\o)+\frac{\kappa}{2}r^2
	-\frac{A_3(\theta)}{3!}r^3
	-\frac{A_4(\theta)}{4!}r^4
	+\Or(r^5).
\end{equation}
Where
\begin{equation}
\eqalign{
A_3(\theta)=-\l \phi(\theta)^3 \r,\\
A_4(\theta)=-\l \phi(\theta)^4 \r +\sum_\alpha \frac{3}{\lambda_\alpha}\l \phi(\theta)^2 \psi_\alpha \r^2.
}
\end{equation}
Then, the condition for stationary point for $\theta$ is
\begin{equation}
\label{eqTheta}
\partial_\theta S_\LS(r,\theta)=0.
\end{equation}
Solving this equation, and substituting the solution $\theta(r)$ into $S_\LS(r,\theta)$
yields one variable function $S_\LS(r,\theta(r))$.
Then, the same arguments for non-degenerated case in \sref{secSufficientCondition}
will be used to describe the sufficient condition for bifurcation.

There may several solutions $\theta(r)$ of equation \eref{eqTheta}.
In such case, each solution yields each $S_\LS(r,\theta(r))$.
As a result, multiple bifurcated solutions corresponds to each solution $\theta(r)$
will emerge from the original at $\kappa=0$.

\subsection{The second derivative of the reduced action}
Before closing this section,
let us consider a meaning of the second derivative of the reduced action.
The second derivative of the reduced action by $r$ is
\begin{equation}
\label{eqSecondDerivativeOfSeff}
S''_\LS(r)=\kappa-A_3 r -\frac{A_4}{2}r^2+\Or(r^3).
\end{equation}
Since this gives the curvature of $S_\LS(r)$ to $r$ direction,
we expect that this expression gives the eigenvalue $\kappa_r$ of the Hessian at $r$,
$\mathcal{H}_r \Phi_r = \kappa_r \Phi_r$,
where
\begin{equation}
\mathcal{H}_r=-\frac{d^2}{dt^2}
	+V_{qq}\left(q+r\phi+r\sum_\alpha \epsilon_\alpha \psi_\alpha\right),\ 
\epsilon_\alpha = -\frac{r}{2\lambda_\alpha}\l \phi^2\psi_\alpha\r +\Or(r^2),
\end{equation}
and  $\Phi_r \to \phi$, $\kappa_r \to \kappa$ for $r \to 0$.
Indeed, in \ref{secEigenvalueForBifurcated},
we will show that 
the eigenvalue $\kappa_r$ to the order $r^2$ is correctly given by \eref{eqSecondDerivativeOfSeff} 
for non-degenerated case.

\section{Properties of third order bifurcation}
\label{secThirdOrder}
\begin{figure}
   \centering
\includegraphics[width=5cm]{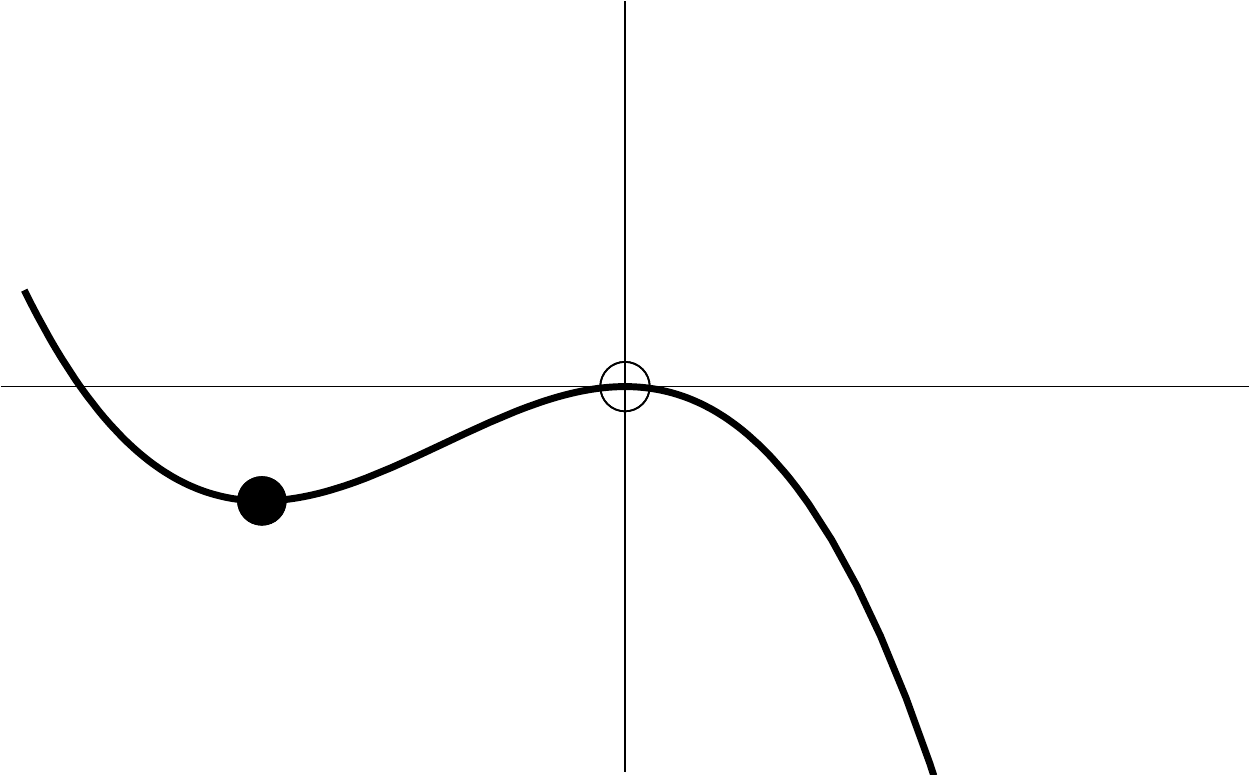}
\includegraphics[width=5cm]{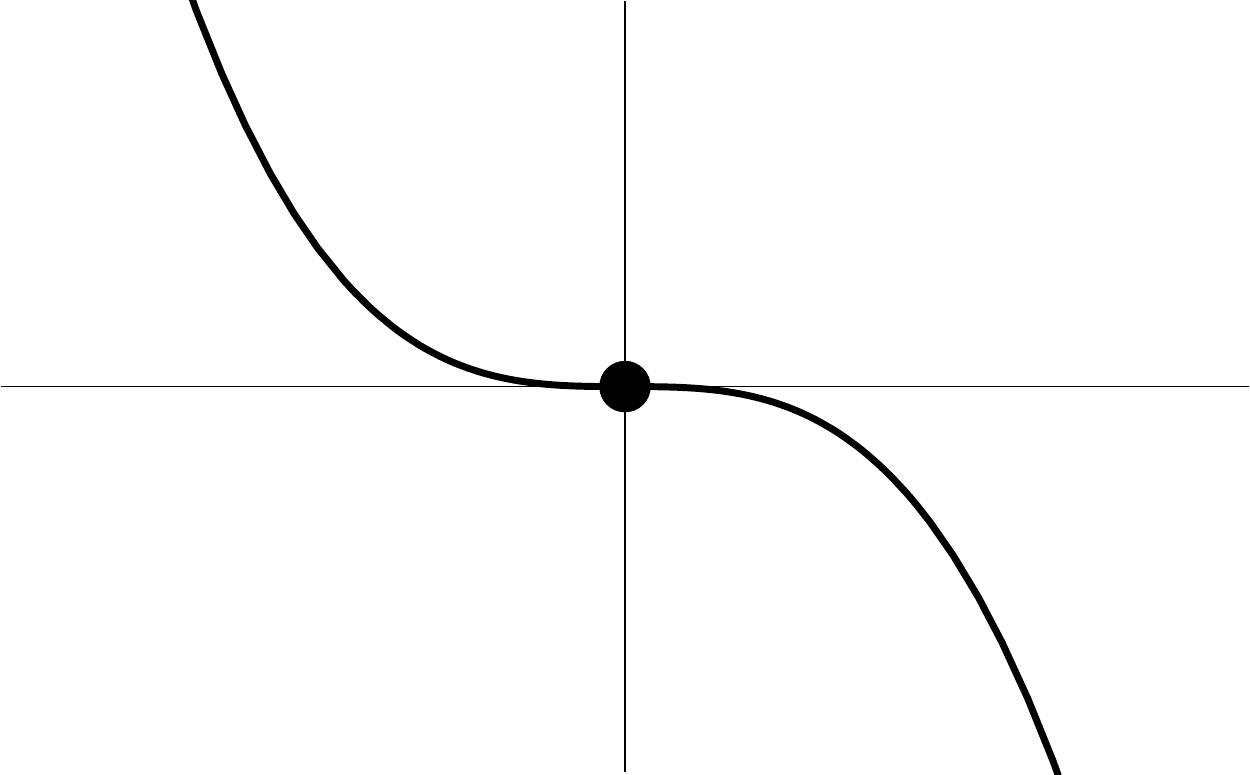}
\includegraphics[width=5cm]{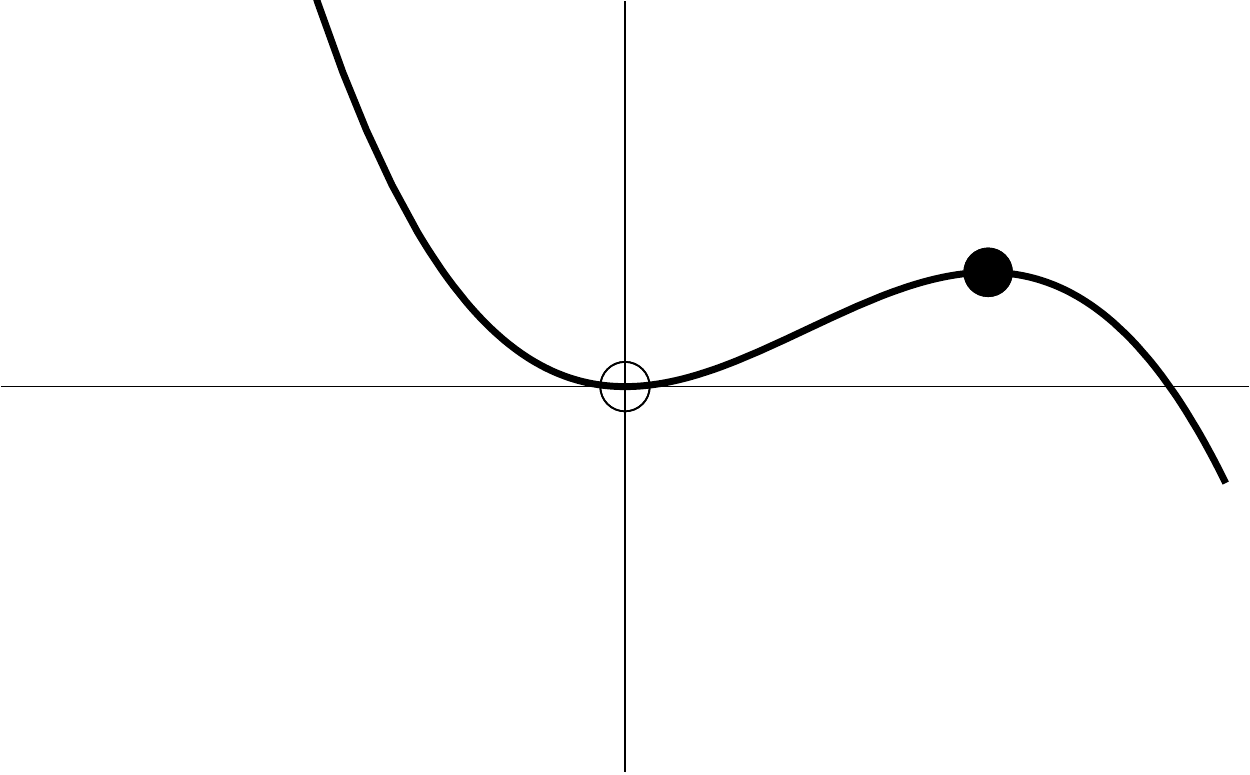}
   \caption{Action truncated to $r^3$, $S_\LS(r)=\kappa r^2/2-A_3 r^3/6$, for $A_3>0$.
   Left to right $\kappa<0$, $\kappa=0$, and $\kappa>0$.
   Hollow and solid circle represents original and bifurcated solution respectively.
   The sign of $S_\LS(r_\b)-S_\LS(r_\o)$ and $\kappa$ are the same.
   The second derivative of this truncated action at $r_\b$ and $r_\o$ satisfies
   $S''(r_\b)=-S''(r_\o)$.
   }
   \label{figThirdOrder}
\end{figure}

If $A_3\ne 0$, a third order bifurcation appears.
The reduced action is
\begin{equation}
\label{SLSforThirdOrder}
S_\LS(r)=\frac{\kappa}{2}r^2-\frac{A_3}{3!}r^3-\frac{A_4}{4!}r^4-\Or(r^5).
\end{equation}
The stationary point of $r\ne 0$ is the solution of
\begin{equation}
\kappa=\frac{A_3}{2}r+\frac{A_4}{6}r^2+\Or(r^3).
\end{equation}
Inversely,
\begin{equation}
\label{eqRforThird}
r=\frac{2}{A_3}\kappa-\frac{4 A_4}{3 A_3^3}\kappa^2+\Or(\kappa^3).
\end{equation}
The distance between $q_\b$ and $q_\o$ is
\begin{equation}
\label{eqDistance}
||q_\b-q_\o||^2
=r^2+\Or(r^4)
=\frac{4}{A_3^2}\kappa^2-\frac{16A_4}{3 A_3^4}\kappa^3+\Or(\kappa^4).
\end{equation}
The difference of the action is given by
\begin{eqnarray}
\label{eqDifferenceOfAction}
\fl
S(q_\b)-S(q_\o)
=\frac{A_3}{2\times 3!}r^3+\frac{A_4}{4!}r^4+\Or(r^5)
=\frac{2}{3 A_3^2}\kappa^3-\frac{2 A_4}{3 A_3^4}\kappa^4+\Or(\kappa^5).
\end{eqnarray}
Since the coefficient for $\kappa^3$ is positive,
the sign of $S(q_\b)-S(q_\o)$ is the same of $\kappa$.
It is obviously shown in \fref{figThirdOrder}.
Combining \eref{eqDistance} and \eref{eqDifferenceOfAction}, we obtain
relation between the difference of the action and the distance,
\begin{equation}
|S_\b-S_\o|
=\frac{|A_3|}{2\times 3!}||q_\b-q_\o||^3
	+\Or\left(||q_\b-q_\o||^4\right).
\end{equation}

The second derivative of the reduced action
$S''_\LS(r)$ at $r=r_\b$ gives
the eigenvalue of the Hessian for the bifurcated solution,
which we call $\kappa_\b$,
\begin{equation}
\label{eqKappaforThird}
\kappa_\b=S''_\LS(r_\b)=-\kappa-\frac{2 A_4}{3 A_3^2}\kappa^2+\Or(\kappa^3).
\end{equation}
The leading term in \eref{eqKappaforThird} is  $-\kappa$
that is independent from coefficients $A_n$.
This is a general property
of the graph $y=ax^2+bx^3$,
which is point symmetric
about the inflection point.
Therefore, the second derivative at two stationary points has opposite sign
and the same magnitude. See  \fref{figThirdOrder}.

In this case, the bifurcated solution exists both side of $\kappa$,
namely, $\kappa>0$ and $\kappa<0$.
The bifurcation point is $\kappa=0$.
This bifurcation is the most simplest case,
however it turns out to describe two bifurcations
that have completely different appearance.
One  is crossing of two solutions. 
The bifurcated solution approaches and crosses the original one at the bifurcation point.
Another one is a pair annihilation or creation of two solutions at the bifurcation point.

In the following subsections, we will explain these two cases.

\subsection{Crossing of two solutions}
Suppose a case where the relation of $\kappa$ and the parameter $\xi$ is given by
\begin{equation}
\kappa = a(\xi-\xi_0)+\Or((\xi-\xi_0)^2),
\end{equation}
where $a$ is a nonzero constant.
Then the correspondence of $\xi$ and $\kappa$ is one to one
in a small region around the bifurcation point $\kappa=0$ or $\xi=\xi_0$.
In this case, the \fref{figThirdOrder} shows $S_\LS(r)$
of the cases $a(\xi-\xi_0)$  negative, zero, and positive.
For both sides of $\xi$,
both the original and bifurcated solution exists.
We will observe the two solution crossing at the bifurcation point.

\subsection{Pair annihilation or creation}
Suppose a case where the relation of $\kappa$ and the parameter $\xi$ is given by
\begin{equation}
\label{eqBehaviourOfXiforAnnihilation}
\xi_0 -\xi= a \kappa^2 +\Or(\kappa^3),
\end{equation}
where $a$ is a nonzero constant.
Let us assume $a>0$ for simplicity.
It is similar for the case $a<0$.
In this case, $\xi$ exists only for $\xi\le \xi_0$.
And there are two values of $\kappa$,
\begin{equation}
\label{eqKappaForPairAnnihilationCreation}
\kappa=\pm \sqrt{a^{-1}(\xi_0-\xi)}\ \left(1+\Or(\xi_0-\xi)\right)
\mbox{ for } \xi<\xi_0.
\end{equation}

\begin{figure}
   \centering
\includegraphics[width=5cm]{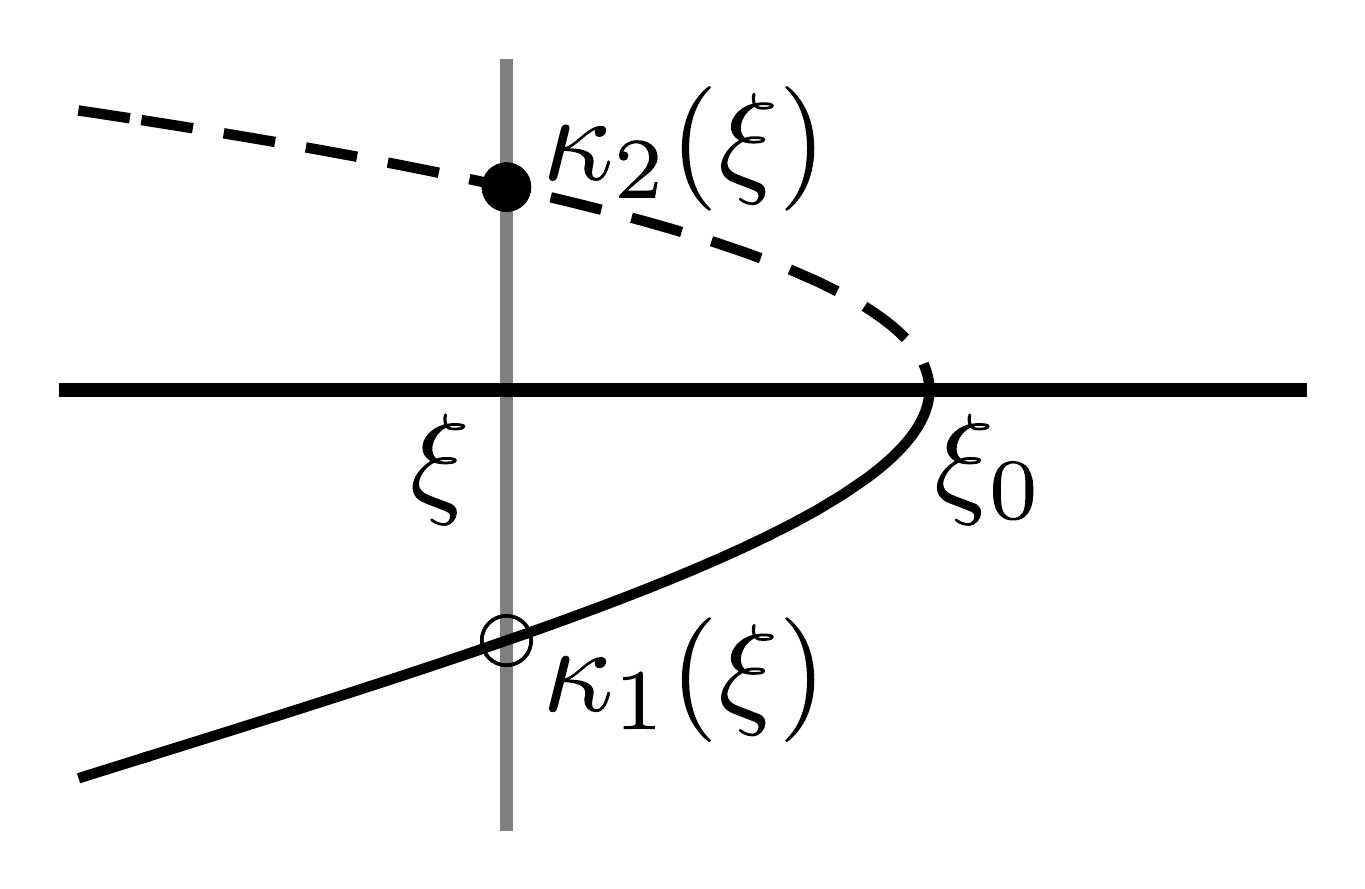}
   \caption{A typical behaviour of parameter $\xi$ and eigenvalue of $\kappa$
   for pair annihilation/creation of solutions $q_1$ and $q_2$.
   Solid and dashed curve represent the eigenvalue for $q_1$ and $q_2$ respectively.
   }
   \label{figXiandKappaForAnnihilation}
\end{figure}

This situation will happen
if there are almost identical two solutions $q_1$ and $q_2$ for parameter in $\xi < \xi_0$.
Approaching $\xi\to \xi_0$, two solutions approach, then vanish at $\xi=\xi_0$.
There are no solutions for $\xi > \xi_0$.
This is a pair annihilation of two solutions at $\xi=\xi_0$.
If we observe this process in inverse direction of $\xi$, this is a pair creation.
A typical behaviour of $\xi$ and $\kappa$ for pair annihilation/creation
is shown in \fref{figXiandKappaForAnnihilation}.

To our knowledge,
we can take alternative parameter $\tilde{\xi}$,
by which we can ``turn around'' the pair  annihilation point smoothly.
The correspondence between $\kappa$ and $\tilde{\xi}$ is locally one-to-one.
So, we can use $\kappa$ for the new parameter $\tilde{\xi}$.
Starting from the solution $q_1$ whose eigenvalue $\kappa_1(\xi)<0$,
changing the parameter $\kappa$ beyond zero,
we will reach $\kappa_2(\xi)>0$. See \fref{figXiandKappaForAnnihilation}.
Therefore, there is no fundamental difference between $q_\o$ and $q_\b$ in this case.
If we take $q_\o=q_1$ at $\kappa=\kappa_1<0$, then $q_\b=q_2$ at $\kappa_2$.
With the same light, we can take $q_\o=q_2$ at $\kappa=\kappa_2>0$, then $q_\b=q_1$.
Assuming the continuity in $\kappa$ of the coefficient $A_n$ in \eref{SLSforThirdOrder},
we can pass through the bifurcation point $\kappa=0$ smoothly
using the parameter $\kappa$.
This is  the situation
that is consistent to our knowledge.

\begin{figure}
   \centering
\includegraphics[width=5cm]{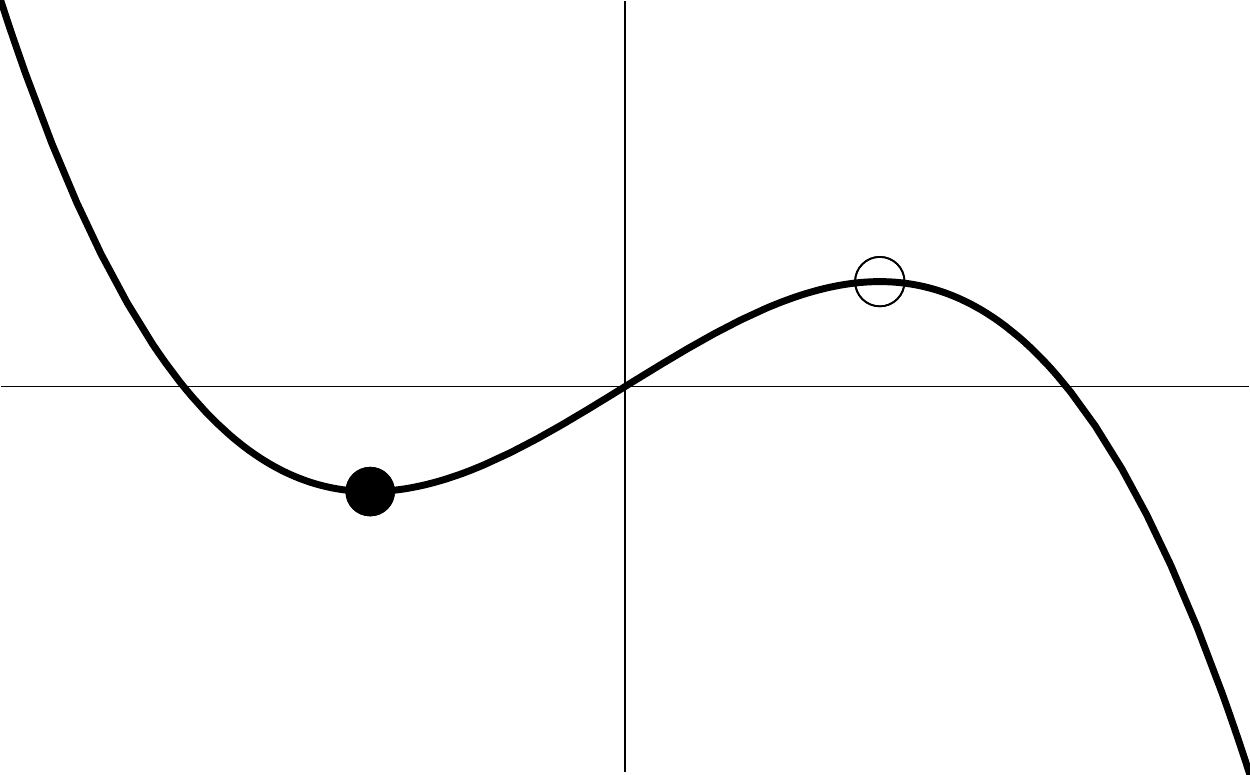}
\includegraphics[width=5cm]{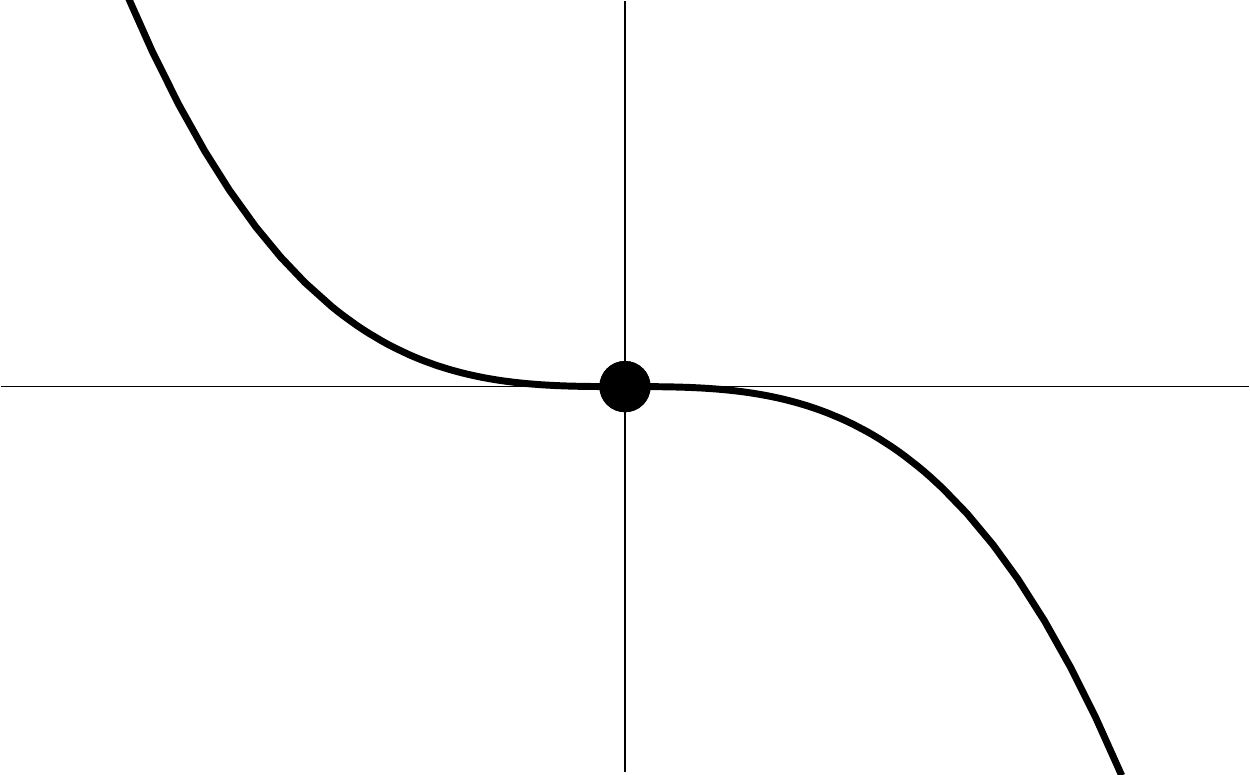}
\includegraphics[width=5cm]{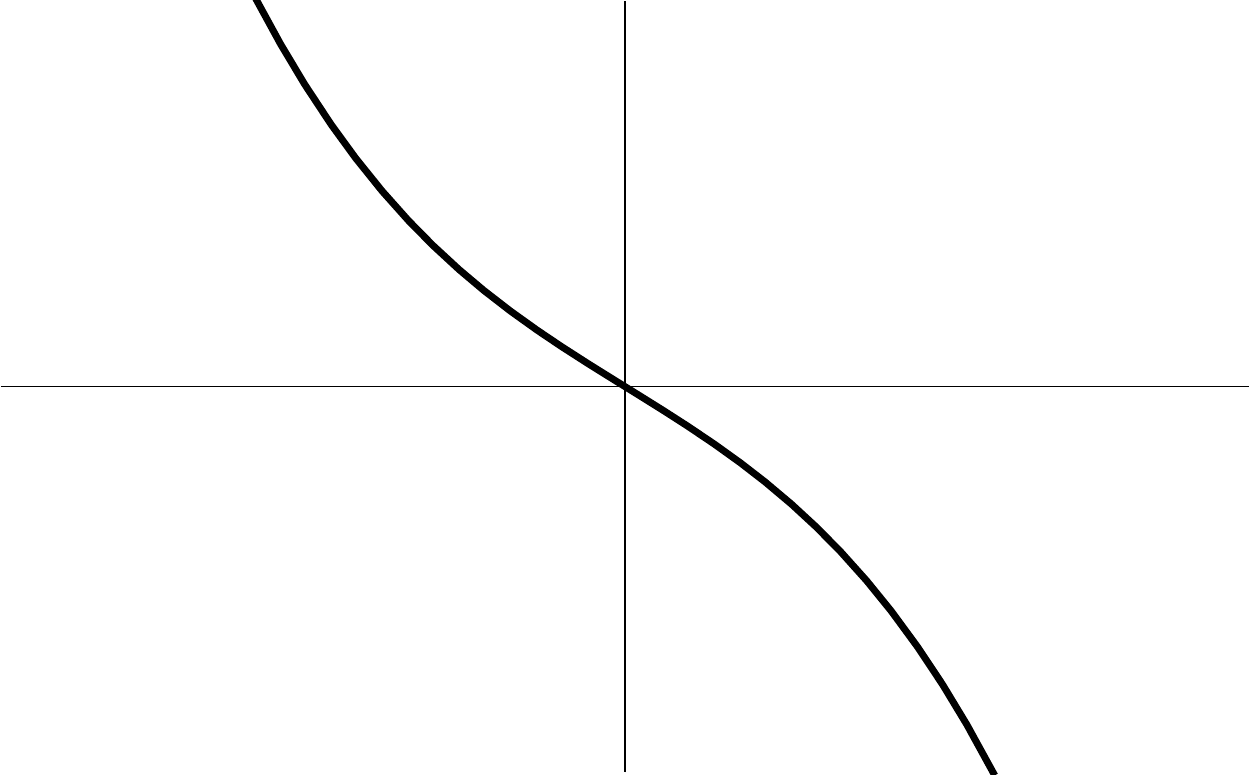}
   \caption{View around an inflection point $r_\I$.
   $S_\LS(r_\I+\tilde{r})=S_\LS(r_\I)+\tilde{A}_1\tilde{r}-\tilde{A}_3\tilde{r}^3/3!$
   for $\tilde{A}_3>0$.
   Left: If $\tilde{A}_1 \tilde{A}_3 >0$, two stationary points exist.
   Middle: One stationary point for $\tilde{A}_1=0$.
   Right: No for $\tilde{A}_1 \tilde{A}_3<0$.
   }
   \label{figViewFromInflectionPoint}
\end{figure}
The situation might be clearly understood by changing the origin of $S_\LS(r)$
to the inflection point $r_\I$ defined by $S''_\LS(r_\I)=0$,
\begin{equation}
r_\I=\frac{\kappa}{A_3}-\frac{A_4}{2 A_3^3}\kappa^2 + \Or(\kappa^3).
\end{equation}
Then
\begin{equation}
\fl
S_\LS(r_\I+\tilde{r})
=S_\LS(r_\I)
	+\left(\frac{\kappa^2}{2 A_3}+\Or(\kappa^3)\right)\tilde{r}
	-\left(\frac{A_3}{6}+\frac{A_4}{6 A_3}\kappa+\Or(\kappa^2)\right)\tilde{r}^3
	+\Or(\tilde{r}^4).
\end{equation}
There are no $\tilde{r}^2$ term, since $r_\I$ is the inflection point.
Note that the sign of the coefficient of $\tilde{r}$ and $\tilde{r}^3$
are determined by the sign of $A_3$ for small $\kappa$.
For both $\kappa>0$ and $\kappa<0$, there are two stationary points.
See \fref{figViewFromInflectionPoint}.
There are no singularity at $\kappa=0$.
While, by the parameter $\xi$,
the behaviour \eref{eqBehaviourOfXiforAnnihilation} is equivalent to 
\begin{equation}
S_\LS(r_\I+\tilde{r})
=S_\LS(r_\I)
	+\left(\frac{a^{-1}}{2 A_3}(\xi_0-\xi)+\dots\right) \tilde{r}
		-\left(\frac{A_3}{3!}+\dots \right) \tilde{r}^3 + \dots.
\end{equation}
Then, if $a>0$,
two stationary points exist for $\xi_0-\xi>0$,
while no  for $\xi_0-\xi<0$.
This describes pair annihilation of solutions.

\section{Properties of fourth order bifurcation}
\label{secFourthOrder}
If there is some reason for $A_3=0$ and $A_4\ne 0$, the reduced action is given by
\begin{equation}
S_\LS(r)=\frac{\kappa}{2}r^2 -\frac{A_4}{4!}r^4 + \Or(r^5).
\end{equation}
The condition for stationary point
\begin{equation}
S'(r)=r\left(\kappa-\frac{A_4}{3!}r^2 + \Or(r^3)\right)=0
\end{equation}
has bifurcated solution 
\begin{equation}
\kappa = \frac{A_4}{3!} r^2+\Or(r^3)
\qquad
\leftrightarrow
\qquad
r=\pm \left(\frac{6\kappa}{A_4}\right)^{1/2}\ (1+\Or(\kappa)).
\end{equation}
In this case, the bifurcated solution exists only  one side of $\kappa$.
\begin{figure}
   \centering
\includegraphics[width=5cm]{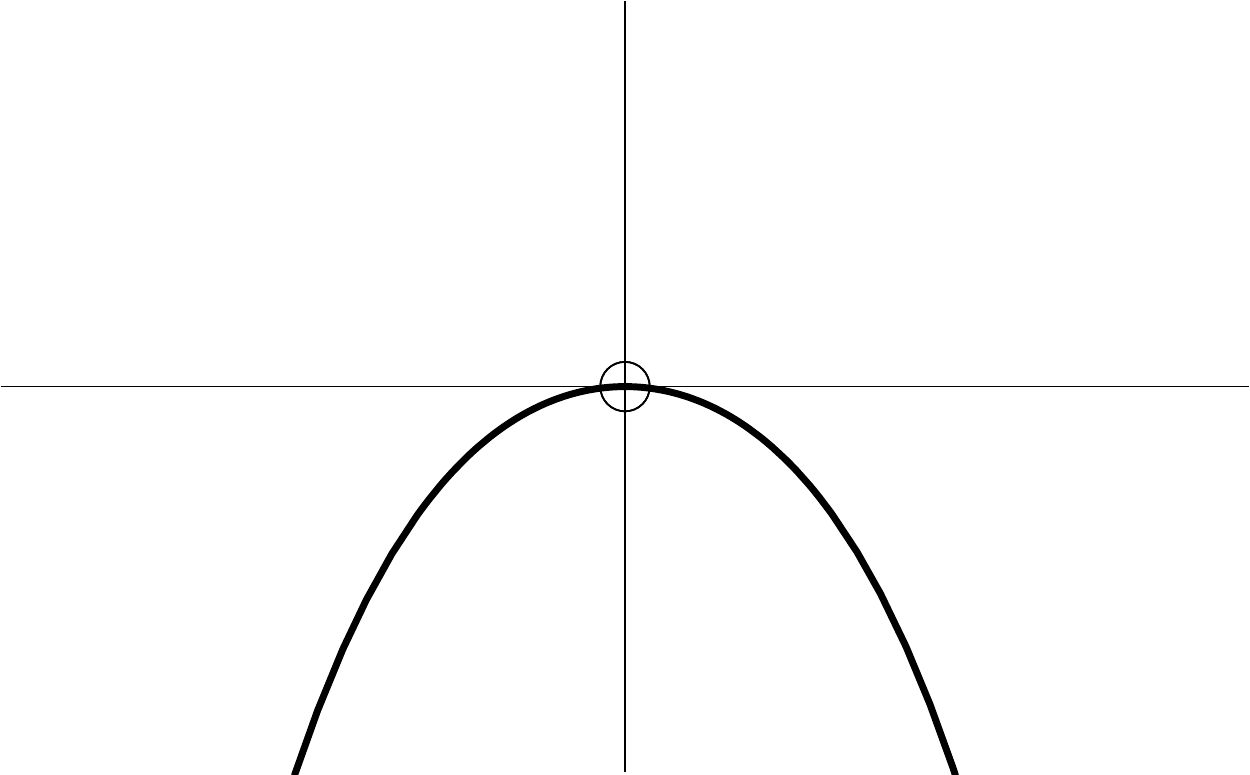}
\includegraphics[width=5cm]{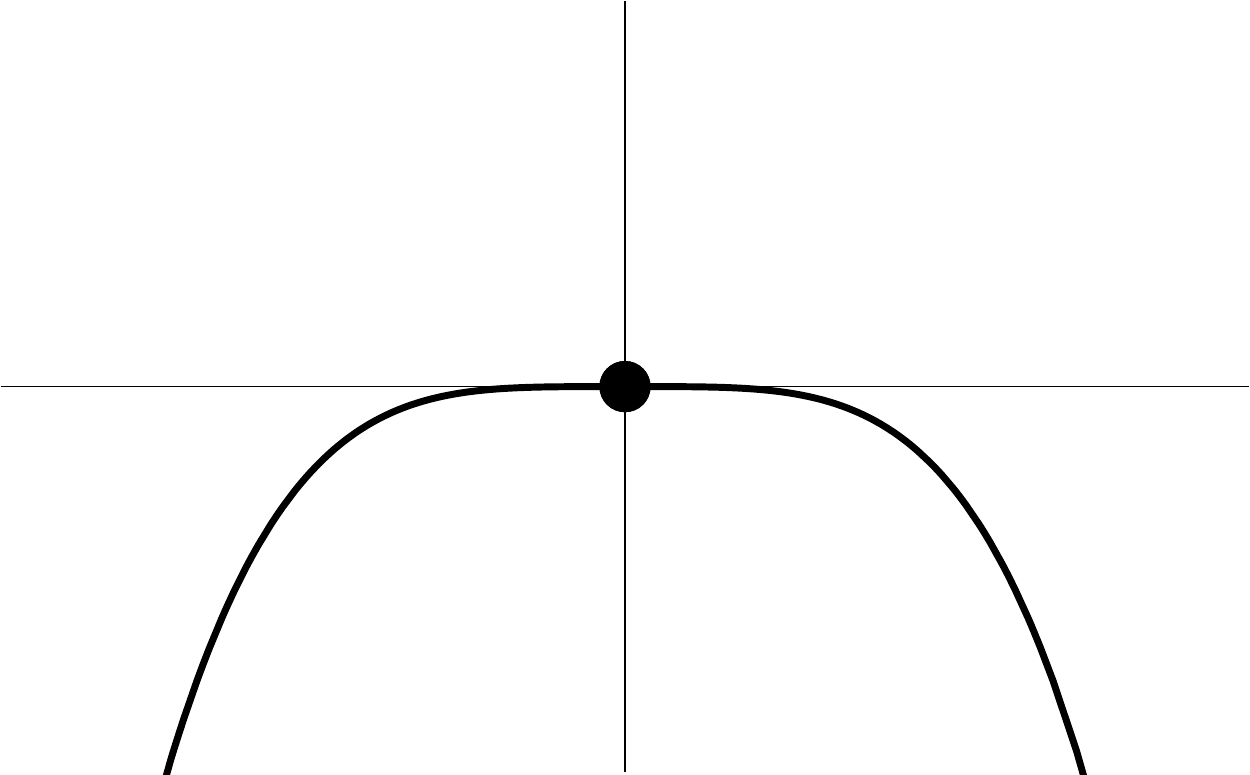}
\includegraphics[width=5cm]{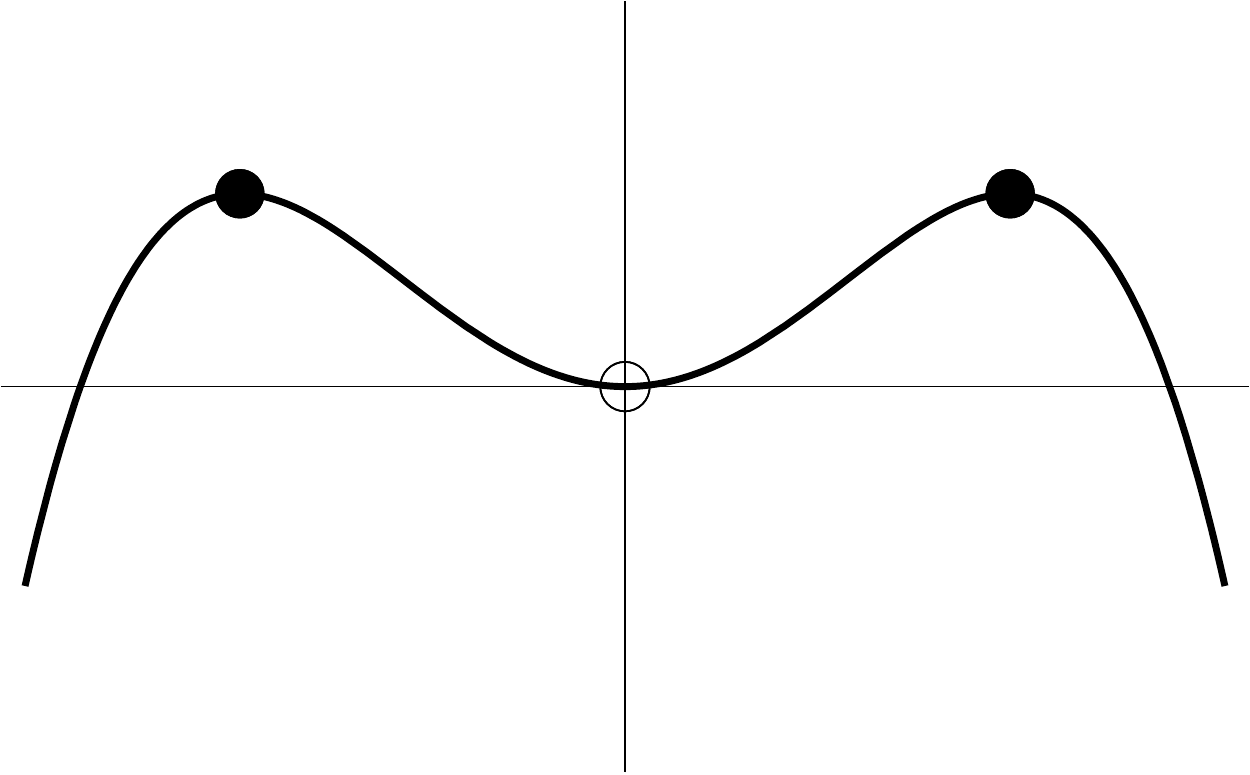}
   \caption{Action truncated to $r^4$, $S_\LS(r)=\kappa r^2/2-A_4 r^4/4!$, for $A_4>0$.
   Left to right $\kappa<0$, $\kappa=0$, and $\kappa>0$.
   Hollow and solid circle represents original and bifurcated solution respectively.
   The sign of $S_\LS(r_\b)-S_\LS(r_\o)$ and $\kappa$ are the same.
   The second derivative of this truncated action at $r_\b$ and $r_\o$
   satisfies $S''(r_\b)=-2 S''(r_\o)$.
   }
   \label{figFourth}
\end{figure}
If $A_4>0$ then it exists in $\kappa>0$, and if $A_4<0$ it exists in $\kappa<0$.
See \fref{figFourth}.
The difference of the action and the distance are
\begin{eqnarray}
||q_\b-q_\o||^2=r^2(1+\Or(r^4))=\frac{6}{A_4}\kappa\left(1+\Or(\kappa)\right),\label{distanceInFourth}\\
S_\b-S_\o
=\frac{A_4}{4!}r^4+\Or(r^5)
=\frac{3}{2 A_4}\kappa^2\left(1+\Or(\kappa^{1/2}\right).\label{actionInFourth}
\end{eqnarray}
Since, the sign of $\kappa$ and $A_4$ is the same,
The sign of $S_\b-S_\o$ is also the same.
This is obviously shown in \fref{figFourth}.
Combining equations \eref{distanceInFourth} and \eref{actionInFourth},
we obtain,
\begin{equation}
S_\b-S_\o
=\frac{A_4}{4!} ||q_\b-q_\o||^4 +\Or(||q_\b-q_\o||^5).
\end{equation}

The second derivative of reduced action with respect to $r$ gives the
eigenvalue of the Hessian at the bifurcated solution,
\begin{equation}
\kappa_\b=S''(q_\b)=-2\kappa\ (1+\Or(\kappa^{1/2}).
\end{equation}
The leading term does not depends on $A_n$.
This is a general property
of the function $y=ax^2+bx^4$.
See \fref{figFourth}.

\section{Summary and discussions}\label{secDiscussions}
An application of variational principle to bifurcation
of periodic solution in Lagrangian mechanic is shown.
The second derivative of the action integral defines linear operator, 
Hessian $\mathcal{H}$.
Eigenvalues of the Hessian has important role for bifurcation.
The necessary and sufficient condition for bifurcation is 
one of the eigenvalue tends to zero. 
For sufficient condition, we couldn't exclude one exceptional case, that unlikely exists.
The eigenvalues of $\mathcal{H}$ being always zero are connected to conservation laws.
Therefore they are irrelevant to bifurcation.
Eigenvalue $\kappa\ne 0$ that tends to  zero yields bifurcation.
The eigenfunction $\phi$ that belonging to the eigenvalue $\kappa$ is
primary for bifurcated solution $q_\b$.

Reduced action $S_\LS$, that is constructed by Lyapunov-Schmidt reduction,
is a function of  one variable or a few variables.
The reduced action $S_\LS(r)$ for non-degenerated $\kappa$
or $S_\LS(r,\theta)$ for degenerated $\kappa$ describes 
the behaviour of the action.
Some derivatives of $S_\LS$ at the original solution $q_\o$
reveal the behaviour of  $S_\LS$ near $q_\o$.
A stationary point of $S_\LS$ is a stationary point of full action,
therefore it is a solution of equations of motion.
Thus we can find a bifurcated solution $q_b$.
The difference of $S_\LS$ between the original and bifurcated solution
gives the difference of full action.
The second derivative of $S_\LS(r)$ at $q_b$ is shown to give
correct eigenvalue of Hessian at $q_b$ to the term $r^2$
for non-degenerated case.
Thus the reduced action $S_\LS$ is important and useful to investigate bifurcations.

Applications for individual Lagrangian mechanics
needs information of symmetry of the solution.
Therefore, they will be published separately. 
Mu\~{n}oz \etal \cite{Munoz2018}, and Fukuda \etal 
\cite{Fukuda2017, fukuda2018, Fukuda2019} found many bifurcations
form figure-eight solutions under homogeneous potential $1/r^a$
or Lennard-Jones potential $1/r^6 - 1/r^{12}$.
The description by $S_\LS$ explains 
qualitative properties of bifurcations and bifurcated solutions
observed by these authors.
Quantitative comparison will be shown in elsewhere.

Period $K$ bifurcations, period doubling etc, will be treated by
relaxing
the boundary condition for variation to $\delta q(t+KT)=\delta q(t)$, $K=2,3,\dots$.
This extends the function space of eigenvalue problem
$\mathcal{H}\Psi=\kappa\Psi$, $\Psi(t+KT)=\Psi(t)$.
Then, an eigenvalue $\kappa \to 0$ gives period $K$ bifurcation.
Some of $5$-slalom solutions found by S\v{u}vakov \etal \cite{Suvakov2013, SuvakovShibayama2016}
can be described as bifurcated solutions of period $5$ bifurcation from the figure-eight.

Equality of the second derivative of $S_\LS(r)$ at bifurcated solution and 
the eigenvalue of Hessian to $r^3$ or higher for non-degenerated case needs further investigation.
The equality for degenerated case will be shown in a separate paper.

Relations between the behaviour of the action and linear stability of 
solutions $q_\o$, $q_\b$
needs further investigations. The relation is unclear at present.

In this paper we use the terms `third order bifurcation', `fourth order bifurcation',
`crossing of solutions' and `pair creation/annihilation of solutions'
instead of `trans-critical', `saddle-node', or `pitchfork'.
The relations between the bifurcations described in this paper
and `trans-critical', `saddle-node', or `pitchfork'
are unclear.

\ack
This work was supported by JSPS Grant-in-Aid for Scientific Research 17K05146 (HF) and 17K05588 (HO).

\appendix
\section{Eigenvalue of Hessian at a bifurcated solution}
\label{secEigenvalueForBifurcated}
In this section, we calculate the eigenvalue of 
\begin{equation}
\mathcal{H}(q_r)=-m\frac{d^2}{d t^2}+V_{q q}(q_r)
\end{equation}
at 
\begin{equation}
q_r=q+r\phi+r\sum \epsilon_\alpha \psi_\alpha$,
$\epsilon_\alpha = -r\l \phi^2\psi_\alpha\r/(2\lambda_\alpha)+\Or(r^2), 
\end{equation}
for non-degenerated case.
Here $\phi$ is the eigenfunction of the Hessian at $q$ with eigenvalue $\kappa$,
\begin{equation}
\label{zerothOrder}
\mathcal{H}(q)\phi=\kappa \phi.
\end{equation}
Expanding $V_{q q}(q_r)$ by $r$, we obtain
\begin{equation}
\eqalign{
\mathcal{H}(q_r)=\mathcal{H}(q)+\Delta V,\\
\Delta V
=\left( r\phi+r\sum \epsilon_\alpha \psi_\alpha\right)V_{q q q}(q)
	+\frac{r^2}{2} \phi^2 V_{q q q q}(q)+\Or(r^3).
}
\end{equation}

We calculate the eigenvalue problem by perturbation theory
with parameter $\rho$,
\begin{equation}
\left(\mathcal{H}(q)+\rho \Delta V\right)\psi = (\kappa+\rho \kappa_1+\rho^2 \kappa_2) \psi
\end{equation}
with the zeroth order \eref{zerothOrder}.
The first order term $\kappa_1$ is
\begin{eqnarray}
\kappa_1=\int_0^T d t\ \phi^2\Delta V 
=r\l \phi^3\r + r \sum \epsilon_\alpha \l \phi^2 \psi_\alpha\r
	+\frac{r^2}{2}\l \phi^4 \r.
\end{eqnarray}
And the second order is
\begin{eqnarray}
\kappa_2
=\sum \frac{1}{\kappa-\lambda_\alpha}\left( \int_0^T d t\ \phi \psi_\alpha \Delta V\right)^2
=\sum \frac{r^2}{\kappa-\lambda_\alpha}\l \phi^2 \psi_\alpha \r^2 + \Or(r^3).
\end{eqnarray}
Substituting the expression for $\epsilon_\alpha$ and using the fact $\kappa=\Or(r)$,
we get the eigenvalue
\begin{equation}
\kappa_r=\kappa
	+r\l \phi^3\r
	+\frac{r^2}{2}\l \phi^4 \r 
	-\frac{3r^2}{2}\sum \frac{1}{\lambda_\alpha}\l \phi^2 \psi_\alpha\r^2 + \Or(r^3),
\end{equation}
which is equal to the expression \eref{eqSecondDerivativeOfSeff}.

\vspace{2pc}
\noindent{\small
{\bf ORCID iDs}\\
Toshiaki Fujiwara: https://orcid.org/0000-0002-6396-3037\\
Hiroshi Fukuda: https://orcid.org/0000-0003-4682-9482\\
Hiroshi Ozaki: https://orcid.org/0000-0002-8744-3968\\
}

\section*{References}
\bibliographystyle{unsrt}
\bibliography{VariationalPrinciple04.bib}

\begin{thebibliography}{10}

\bibitem{Landau}
L.~D. Landau and E.~M. Lifshitz.
\newblock {\em Mechanics}, volume~1 of {\em Course of Theoretical Physics}.

\bibitem{Arnold}
V.~I. Arnold.
\newblock {\em Mathematical Methods of Classical Mechanics}, volume~60 of {\em
  Graduate Texts in Mathematics}.

\bibitem{Moore}
Cristopher Moore.
\newblock Braids in classical dynamics.
\newblock {\em Phys. Rev. Lett.}, 70:3675--3679, Jun 1993.

\bibitem{ChencinerMontgomery}
Alain Chenciner and Richard Montgomery.
\newblock A remarkable periodic solution of the three-body problem in the case
  of equal masses.
\newblock {\em Annals of Mathematics}, 152(3):881--901, 11 2000.

\bibitem{RotatingEight}
Alain Chenciner, Jacques F{\'e}joz, and Richard Montgomery.
\newblock {Rotating Eights: I. The three {$\Gamma_i$} families}.
\newblock {\em Nonlinearity}, 18:1407, 03 2005.

\bibitem{GuoWu2013}
Shangjiang Guo and Jianhong Wu.
\newblock {\em Bifurcation Theory of Functional Differential Equations}, volume
  184 of {\em Applied Mathematical Sciences}.
\newblock Springer, 2013.

\bibitem{Munoz2018}
Francisco~Javier Mu{\~n}oz-Almaraz, Jorge Gal{\'a}n-Vioque, Emilio Freire, and
  Andre Vanderbauwhede.
\newblock {Numerical explorations in a modified potential of the TBP}.
\newblock https://doi.org/10.5281/zenodo.1500051, November 2018.

\bibitem{Fukuda2017}
Hiroshi Fukuda, Toshiaki Fujiwara, and Hiroshi Ozaki.
\newblock Figure-eight choreographies of the equal mass three-body problem with
  {Lennard-Jones-type} potentials.
\newblock {\em Journal of Physics A: Mathematical and Theoretical},
  50(10):105202, Feb 2017.

\bibitem{fukuda2018}
Hiroshi Fukuda, Toshiaki Fujiwara, and Hiroshi Ozaki.
\newblock Morse index for figure-eight choreographies of the planar equal mass
  three-body problem.
\newblock {\em Journal of Physics A: Mathematical and Theoretical},
  51(14):145201, Mar 2018.

\bibitem{Fukuda2019}
Hiroshi Fukuda, Toshiaki Fujiwara, and Hiroshi Ozaki.
\newblock Morse index and bifurcation for figure-eight choreographies of the
  equal mass three-body problem.
\newblock {\em Journal of Physics A: Mathematical and Theoretical},
  52(18):185201, Apr 2019.

\bibitem{Suvakov2013}
Milovan {\v{S}}uvakov and V.~Dmitra\ifmmode \check{s}\else
  \v{s}\fi{}inovi\ifmmode~\acute{c}\else \'{c}\fi{}.
\newblock Three classes of newtonian three-body planar periodic orbits.
\newblock {\em Phys. Rev. Lett.}, 110:114301, Mar 2013.

\bibitem{SuvakovShibayama2016}
Milovan {\v{S}}uvakov and Mitsuru Shibayama.
\newblock Three topologically nontrivial choreographic motions of three bodies.
\newblock {\em Celestial Mechanics and Dynamical Astronomy}, 124(2):155--162,
  Feb 2016.

\end{thebibliography}

\end{document}